\documentclass{sig-alternate}
\newcommand{\ignore}[1]{}
\usepackage[pass]{geometry}
\usepackage{fancyhdr}
\usepackage[normalem]{ulem}
\usepackage[hyphens]{url}
\usepackage{hyperref}

\usepackage{graphicx}
\usepackage{epstopdf}
\usepackage{multirow}
\usepackage{amsmath}
\usepackage{algorithm}
\usepackage[noend]{algpseudocode}
\usepackage{authblk}

\algrenewcommand\algorithmicindent{1.0em}%

\algnewcommand{\Initialize}[1]{%
  \State \textbf{Initialize:}
  \Statex \hspace*{\algorithmicindent}\parbox[t]{.8\linewidth}{\raggedright #1}
}
\newcommand{\proposedarch}{NeuroTrainer}

\fancypagestyle{firstpage}{
  \fancyhf{}
\setlength{\headheight}{50pt}

  \pagenumbering{arabic}
}

\title{\proposedarch: An Intelligent Memory Module for Deep Learning Training} 
\author[1]{Duckhwan Kim}
\author[1]{Taesik Na}
\author[1]{Sudhakar Yalamanchili }
\author[1]{Saibal Mukhopadhyay}

\affil[1]{\footnotesize School of Electrical and Computer Engineering, Georgia Institute of Technology, Atlanta, GA 30332, USA}

\begin{document}
\maketitle

\thispagestyle{firstpage}
\pagestyle{plain}

%%%%%% -- PAPER CONTENT STARTS-- %%%%%%%%
\begin{abstract}
This paper presents,~\proposedarch, an intelligent memory module with in-memory accelerators that forms the building block of a scalable architecture for energy efficient training for deep neural networks. The proposed architecture is based on integration of a homogeneous computing substrate composed of multiple processing engines in the logic layer of a 3D memory module. NeuroTrainer utilizes a programmable data flow based execution model to optimize memory mapping and data re-use during different phases of training operation. A programming model and supporting architecture utilizes the flexible data flow to efficiently accelerate training of various types of DNNs. The cycle level simulation and synthesized design in 15nm FinFET shows power efficiency of ~500 GFLOPS/W, and almost similar throughput for a wide range of DNNs including convolutional, recurrent, multi-layer-perceptron, and mixed (CNN+RNN) networks.  

% This paper shows processor in memory accelerator for deep learning training by covering all three phases: feedforward, backpropagation, and weight update with different precision modes for high accuracy. As an example, hybrid memory cube (HMC) with 16 vaults is selected as 3D stacked memory and 15 processing elements (PE) are placed on the logic die of HMC. By separating data path for common data shared by all PEs and independent data for each PE, it can reduce data congestion significantly. The main arithmetic unit, multiplication and accumulation (MAC), can compute two pairs of 16 bit fixed point operands for inference or one pair of 32 bit fixed point with stochastic rounding for training to maintain high accuracy. Proposed architecture supports different data movements and operating modes by layer-wise programming for different required operations in training CNN, RNN, and more complex DNN and it achieves up to 4.7 TOPS/s. In this paper, programming model for address generator (PMAG) and processing elements for each operation is illustrated in detail and hardware implementation using 15nm FinFet is studied in terms of power, area, and thermal issue.
\end{abstract}

%%%% TODO 2017. 7. 28

%% 2. Layout (P&R)

%% 4. Introduction: focusing on why training accelerator is required. 
%% 5. Preliminaries: key architecture challenges in training and shows 'training' accelerator can also do 'inference'
%% 6. 

\section{Introduction}

\begin{figure}[t]
  \centering
  \includegraphics[width=1\columnwidth]{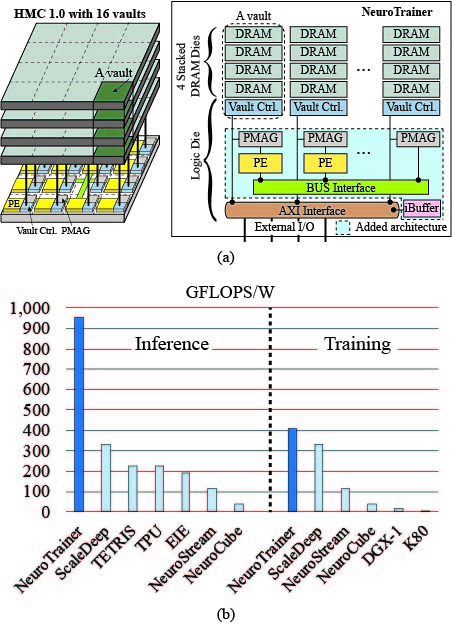}
  \caption{NeuroTrainer overview: (a) Proposed architecture. (b) Efficiency (GFLOPS/W)} 
  \label{fig:training_efficiency}
\end{figure}

The hardware acceleration of \textit{inference} of Deep Neural Network (DNNs) including convolutional (CNN), recurrent (RNN), and multi-layer-perceptron (MLP) have received
considerable attention in recent past ~\cite{han2016eie,chen2014dadiannao,chen2016eyeriss,kim2016neurocube,gao2017tetris,
  azarkhish2017neurostream,nurvitadhi2016accelerating,shin201714,TPU}.  In contrast, training has largely been accelerated
by software implementations executing on clusters of graphics
processing units (GPUs). As DNNs become larger and more
complex, the time and energy costs of training become limiters to the
application of DNNs to more complex problems. For example, a DGX-1
with 8 GPUs consumes more than 3KW, and is limited to training only
1,000 Imagenet images per second for VGG
16~\cite{tensorflow_performance}.  Hence, availability of a
specialized modular architecture for energy efficient scaling of
training performance will be critical to the feasibility of future,
large scale DNNs.

The training of a DNN is composed of three primary steps: forward propagation (FP), which is identical to inference, back propagation (BP), and parameter update (UP) (see Section~\ref{pre} for details). The  acceleration of training faces major additional challenges over acceleration of inference as discussed below. 
%\begin{itemized}

1) Most DNN accelerators for inference are optimized for convolutions
with small kernels and matrix-matrix multiplication (for fully connected layers).
However, accelerating BP and UP includes the following additional
operations - i) convolution with very large kernels, ii) matrix
transpose, iii) vector to vector outer product, iv) loss function
computation, v) a pooling layer and its derivative, and vi) the
derivative of non-linear activation functions.. 

2) Training operates over very large data sets and employs mini-batch processing across training thereby
  requiring larger on-chip storage to increase effective memory bandwidth. In contrast, inference operates over single
  sample. Further, while inference requires only reading weights, training requires reading/writing weights and their gradients, increasing memory traffic.

3) The computation of gradients in back propagation and weight update
require higher bit precision to account for small gradient values (vanishing gradient issue~\cite{hochreiter1991untersuchungen,hochreiter2001gradient}). Therefore, low bit
precision (8/16 bit) arithmetic, often used during inference for
energy efficiency, is not suitable for training.  
%\end{itemized}

% As a consequence of the preceding, training and inference have had
% separate optimized implementations. However, we observe that
% inference and training do share many common features in terms of data
% flow characteristics, arithmetic patterns (e.g.,
% multiply-accummulate), fine-grain concurrency, and memory level
% parallelism. These common features form the nucleus of a programmable
% data flow execution model and in-memory accelerator architecture for
% high performance energy efficient training as illustrated in
% Fig.~\ref{fig:training_efficiency}. 

% There are few recent reports of hardware accelerators that shows the
% potential of higher training-efficiency over GPUs
% (Fig.~\ref{fig:training_efficiency}). Neurocube~\cite{kim2016neurocube}
% and Neurosteam \cite{azarkhish2017neurostream} showed the use of
% in-memory acceleration can enhance energy-efficiency of training;
% however, did not present any training specific architectural
% considerations. More recently,
% Scaledeep~\cite{venkataramani2017scaledeep} presented a efficient
% training accelerator composed of a multi-chip module where different
% chips are optimized for different computational kernel(s). However, as
% composition of the multi-chip module is decided during design, given a
% design the training performance can vary significantly for different
% DNNs with varying numbers of convolution, fully connected, and
% recurrent layers. 

This paper presents,~\proposedarch, an intelligent memory module with in-memory accelerators that forms
the building block of a scalable architecture for energy efficient training. The key distinguishing feature of
NeuroTrainer is its programmable data flow execution model. We observed that distinct computational kernels in training different networks share common arithmetic operations (e.g., multiply-and-accumulate) but differ in their memory usage and data flow pattern. Hence,~\proposedarch~utilizes a homogeneous architecture but with an execution model where memory mapping, re-use, and data flow for different kernels are programmed to match the data usage/flow pattern, and hence, optimize performance of each kernel. It is in contrast to recent report of multi-chip module based training accelerator where each chip is independently optimized for a specific operation, creating a heterogeneous architecture~\cite{venkataramani2017scaledeep}. A major advantage of the programmable data-flow based execution on homogeneous substrate, compared to customized hardware based solutions is to provide stable performance over many applications.

The NeuroTrainer is evaluated using cycle level simulation, and synthesized in 15nm FinFET technology. The NeuroTrainer demonstrates 25x higher efficiency over GPU (Fig.~\ref{fig:training_efficiency}).
Moreover,~\proposedarch~ shows higher average power efficiency over prior training accelerators(\cite{kim2016neurocube,azarkhish2017neurostream, venkataramani2017scaledeep}), and more importantly, demonstrate ability to train different types of benchmark networks (CNN, RNN, and CNN + RNN) with similar power-efficiency. We further demonstrate scalable system  with multiple interconnected NeuroTrainers to scale training performance for very large DNNs. Hence, NeuroTrainer can be used as the building block to design a large scale DNN training platform.  

This paper makes the following contributions. 
%\begin{enumerate}

1) We present a programmable data flow based execution model that enables the use of a homogeneous computing architecture to efficiently train diverse DNNs. This flexible data flow programmability enables efficiently accelerate
training of various types of DNNs including CNNs, RNNs, MLPs and hybrid networks (CNN+RNN).

2) We present the~\proposedarch~as a 3D memory module with an integrated in-memory accelerator. The architecture of the memory module is patterned after the Hybrid Memory Cube (HMC) which is composed of stacked DRAM partitioned across multiple independently controlled \textit{vaults}. Each vault has an independent vault controller on the logic die; therefore multiple partitions in a DRAM die can be accessed simultaneously. 

3) We present an in-memory accelerator composed of an array of interconnected processing engines (PEs), implemented on the logic die with precision-configurable arithmetic and support for dataflow based execution. All but one memory vault are connected to a dedicated PE; and the remaining vault is connected to all the PEs using a shared bus. Each vault controller is augmented with a
Programmable Memory Address Generator (PMAGs). 

4) We develop programming model for the~\proposedarch. Compilation now
involves optimized mappings of data (input, parameters, and gradients) into memory vaults, and programming
the PMAGs to orchestrate an efficient flow of data between the DRAM and logic layer in a manner that optimizes bandwidth, exploits re-use, maximizes concurrency, and minimizes data movement and buffering in
the PEs. 

The rest of the paper is organized as follows. Section~\ref{pre} introduces DNN training; Section~\ref{arch} illustrates the proposed architecture; Section~\ref{programing} explains the programming model; Section~\ref{simul} presents simulation results, followed by related work and conclusion.

\section{Preliminaries} \label{pre}

\begin{figure*}[t]
  \centering
  \includegraphics[width=1\textwidth]{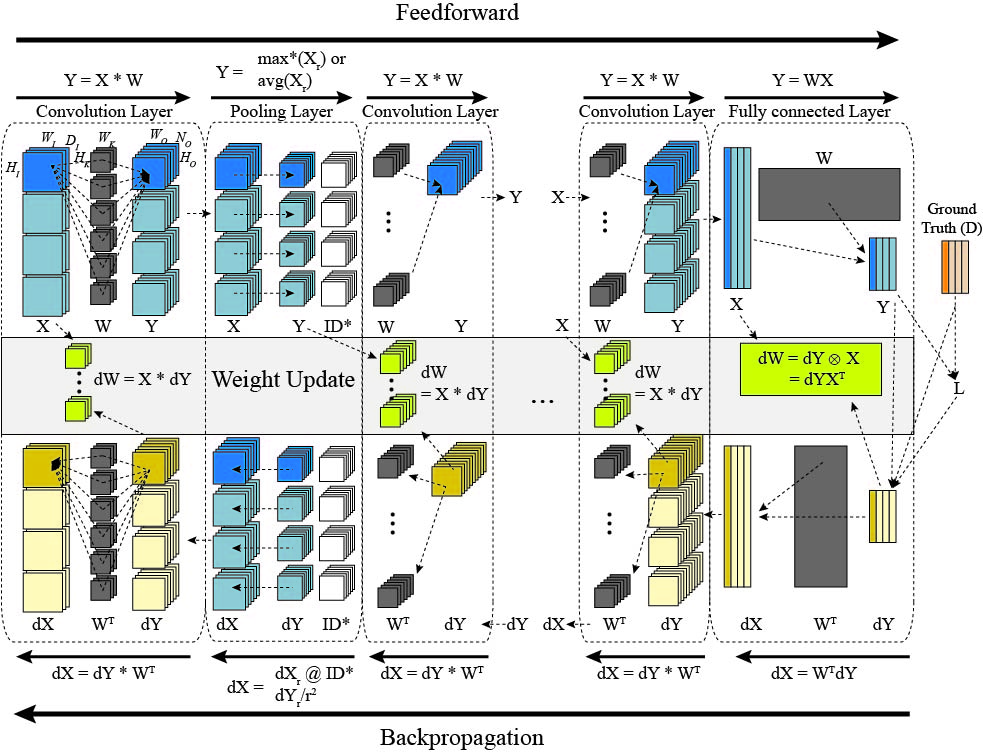}
  \caption{Deep neural network composed of convolution layer, pooling layer, and fully connected layer. It has three phases: feedforward, backpropagation, and weight update (ID*: location of maximum pixel in pooling area).  } 
  \label{fig:op_elem}
\end{figure*}

%\subsection{Training deep neural network with gradient descent} \label{pre:train}
In this section, we will explain the approach for training DNN with gradient descent, which is composed of three steps: feedforward, backpropagation, and weight update in recent DNNs~\cite{haykin2009neural}. Fig.~\ref{fig:op_elem} shows a simple DNN and its feedforward, backpropagation, and weight updating for different types of the layer in the network.

% \begin{figure}[t]
%   \centering
%   \includegraphics[width=1\columnwidth]{./operating_elements_short.eps}
%   \caption{Feedforward (FF), Backpropagation (BP), and Weight Update (UP) for common layers in recent DNNs: convolution layer. ($*$ represents convolution). Underlined input is common input for multiple PEs' paralell operation.} 
%   \label{fig:op_elem}
% \end{figure}

\subsection{Feedforward (FF)}
Feedforward is propagation of neuron activation from $i^{th}$ layer to $i+1^{th}$ layer through weights between two layers. The output of a neuron (state) is weighted summation of activation from all connected neurons in previous layer (and current layer as well for RNN~\cite{elman1990finding}). It is the only phase required during the inference. Fig.~\ref{fig:op_elem} shows feedforward through convolution layer and fully connected layer. 

\subsection{Backpropagation (BP)}
For a given input, at the end of feedforward operation, the output of the last layer is compared with the \textit{ground truth} i.e. the desired output for this input and computing loss ($L$). The loss can be defined by simple mean squared error (MSE) or combination of softmax layer and cross-entropy layer~\cite{dunne1997pairing}. Backpropagation is the phase to find the impact of each state on the loss (gradient) $\partial L / \partial X (dX)$ by propagating from the last layer. Since there is no definition of loss ($L$) in hidden layer, $\partial L / \partial X (dX)$ is computed from the $dX$ of $i+1^{th}$layer ($\partial L / \partial Y (dY)$) instead of computing $\partial L / \partial X (dX)$ directly. We can see most of \textit{arithmetic} operation in Backpropagation is similar to that of Feedforward in convolution and fully connected layer except \textit{transposing kernel} (Fig.~\ref{fig:op_elem}). 

\subsection{Weight Updates (UP)}
Based on $dX$, $\partial L / \partial W (dW)$ needs to be generated to reduce $L$ in next iteration (epoch). New W for next iteration is determined as $W_{new} = W_{old} - \eta \cdot dW$, where $\eta$ is learning rate. Recently, additional term is added during the update such as momentum~\cite{qian1999momentum}. For convolution, it needs convolution between $X$ and $dY$. As dimension of $dY$ is smaller than $X$ by radius of kernel ($W$), it is convolution with very large kernel size. For fully connected layer, it requires vector vector outer product. Thus we can see there is \textit{additional} operation and data flow is needed for efficient operation in weight updating.

\subsection{Data Preparation (Prep)}
For each operation, input data need to be pre-loaded into the memory to improve data flow between memory and processing engines. If the layout of output generated in layer $i$ does not match with required data layout for the layer $i+1$, it needs to be re-arranged between multiple memory banks. In addition, for convolution, to make the size of the output same as the size of input, the input needs \textit{dummy} zeros on its boundary. 

\subsection{Minibatch Training}
% If weights are updated for every sample in training data, it's stochastic gradient descent (SGD)~\cite{haykin2009neural}. It's fast but has a lot of overshoot (risk of stuck at local minimum). If weights are updated after training entire training dataset, it's batch gradient descent (GD). It shows stable cost reduction, but it's slow~\cite{haykin2009neural}. 

Mini batch training involves updating weights after training small set ($K$) of training data. If total number of training data is $N$, it will iterate $N/K$ times for one epoch. It is faster than training with large batch sizes and shows smoother convergence than training individual images. Moreover, it can reuse weights $K$ times improving computing efficiency~\cite{haykin2009neural}. However, it requires more on-chip memory to store $K$ temporal data. 

The multiple mini-batches are trained in parallel using multiple computing nodes where each node independently compute $\partial L / \partial W (dW)$. After generating all $dW$s, all machine share updated new $W$ (synchronous training)~\cite{iandola2016firecaffe}. To overcome the unbalance in training latency among multiple nodes, in asynchronous training once a node generates $dW$, it will have new $W$ while others use old $W$~\cite{dean2012large}.

\section{Proposed Architecture} \label{arch}

% \begin{figure}[t]
%   \centering
%   \includegraphics[width=1\columnwidth]{./Arch_Overview.eps}
%   \caption{Proposed architecture as processor in memory within a Hybrid Memory Cube~\cite{hmc}.} 
%   \label{fig:arch_overview}
% \end{figure}

%\subsection{Processor in Memory (PIM)}
%Low operation density (Ops/Byte) is well known challenge in Deep Learning accelerator design. To reduce overhead to acceess large capacity off-chip memory (relying on only on-chip memory is not enough for recent DNN), placing processing elements (PEs) below stacked memory dies (PIM: processor in memory) has been introduced~\cite{kim2016neurocube,azarkhish2017neurostream,gao2017tetris}. In hybrid memory cube (HMC) with 16 vaults (each vault's data bandwidth is 10GByte/s), \proposedarch is placed on the logic die as illustrated in Fig.~\ref{fig:training_efficiency}.

% As illustrated in Figure~\ref{fig:training_efficiency}, the
% NeuroTrainer logic layer is comprised of an interconnected network of
% PEs, where each PE contains clusters of computation
% units and local buffers for inputs (states) and parameters. Each PE
% has a one high bandwidth connection to its local vault and an
% interface to a broadcast bus connected to a shared vault. Partioning and mapping of the weights
% and states across the vaults will determine the data flow necessary for
% execution. NeuroTrainer adopts a hybrid data flow scheme described in
% the following.

In this paper, \proposedarch~is designed considering a Hybrid Memory Cube (HMC) where a 3D memory stack is partitioned into multiple parallel vaults. For example, HMC 1.0 is composed of 4 DRAM dies partitioned into 16 vaults, each vault has an independent memory controller (vault controller, VC), and connected to 4 external (off-chip) links via AXI interface. The computation fabric of ~\proposedarch~is composed of multiple processing engines (PEs) where each PE contains clusters of computation units and local buffers for inputs (states) and parameters. Each PE has a one high bandwidth connection to its local vault and an
interface to a broadcast bus connected to a shared vault. The vault controllers are augmented with a programmable memory address generator (PMAG), a state-machine that realizes mapping of the different types of data (input, parameter, and gradients) to different vaults and control data flow between memory and PEs. 

%.with all the PEs. There are two separate data paths between PEs and memory: (1) direct communication between a PE and a vault and (2) BUS interface for broadcasting from a vault to all PEs or merging data from all PEs into a single vault. 
%In contrast to~\cite{kim2016neurocube}, which encapsulates the data into packet for network-on-chip (NoC) operation, ~\proposedarch eliminates the overheads of NoC, router, and packet based communication, thereby increasing energy-efficiency.

%~\proposedarch~operation is controlled by data flow without any global controller. 
% From the vault, data is pushed to either PE or BUS, and the PE operates when two input buffers are filled (ready). The vault controllers are augmented with a programmable memory address generator (PMAG), a state-machine that realizes mapping of the different types of data (input, parameter, and gradients) to different vaults and control data flow between memory and PEs. 

%The programming is performed using a small on-chip table placed in the logic layer, referred to as the instruction buffer (iBuffer), that stores the parameters for PMAG to realize different layers of a DNN. Given a DNN to train, the host first programs the iBuffer. Once iBuffer is programmed, the ~\proposedarch operates autonomously and during execution of individual layers each PMAG receives the corresponding table entries from iBuffer. 

\subsection{Hybrid Data Flow} \label{subsec:hybrid_data_flow}

\begin{figure}[t]
  \centering
  \includegraphics[width=1\columnwidth]{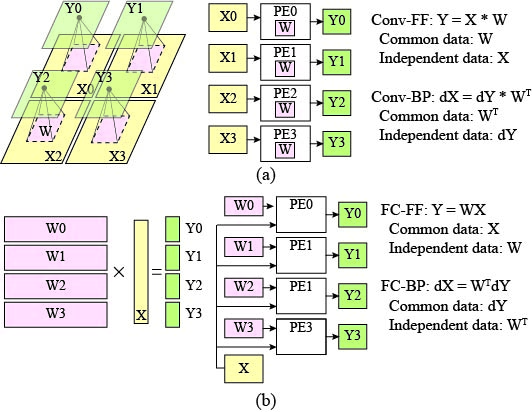}
  \caption{Hybrid Data Flow. (a) 2D convolution with small kernels and (b) Matrix multiplication} 
    % \caption{Hybrid Data Flow. (a) 2D convolution with small kernels. Small common data (kernels: W) is pre-stored in PE's memory. (b) Matrix multiplication. Large common data (X) is broadcast to all PEs and partial weight matrix (W0 - W3) is allocated into 4 PEs in parallel.} 
  \label{fig:hybrid_data_flow}
\end{figure}

% We distinguish between small common data and large common data. For
% example, the weights of small kernels in the convolution layer would
% be small common kernels whereas the large weight matrix in a fully
% connected layer corresponds to large common data. The approach used in
% NeuroTrainer is to buffer copies of small common data across PEs
% and stream partitions of large data (e.g. inputs to a layer) from the local vault. Alternatively,
% with respect to large common weight matrices, they can be
% broadcast from a shared vault to all PEs, while partial weight matrices
% are stored across PEs. These two classes of data flows are illustrated
% in Figure~\ref{fig:hybrid_data_flow}.

% \proposedarch~is designed to provide two different data movements between vaults and PEs based on the operation types. From Fig.~\ref{fig:op_elem}, the most important operations in DNN are convolution and matrix-matrix multiplication. During the convolution (Fig.~\ref{fig:hybrid_data_flow} (a)), kernel ($W$) is shared by 4 PEs for different partial outputs ($Y0 = X0 * W$, ... , $Y3 = X3 * W$) while input is partitioned for each PE ($X0 \sim X3$). 
% For the matrix-matrix multiplication, weight matrix ($W$) is partitioned to 4 blocks row-wise ($W0 \sim W3$) while input ($X$) is shared by 4 PEs for different partial outputs ($Y0 = W \cdot X0$, ... ,$Y3 = W \cdot X3$). We should note that one of the inputs can be shared by all PEs in any operations of DNN (underlined input in Fig.~\ref{fig:op_elem}).

\proposedarch~is designed to provide two different data movements between vaults and PEs based on the operation types. Consider convolution and matrix-matrix multiplication. During the convolution (Fig.~\ref{fig:hybrid_data_flow} (a)), kernel ($W$) is shared by PEs while input is partitioned for each PE. For the matrix-matrix multiplication, weight matrix ($W$) is partitioned while input ($X$) is shared by PEs. We note that one of the inputs can be shared by all PEs in any operations of DNN.

Based on the size of common input, operations in Fig.~\ref{fig:op_elem} can be classified as \textit{small common data} (ex: convolution with small kernels) and \textit{large common data} (ex: matrix-matrix multiplication). For example, the weights of small kernels in the convolution layer would
be small common kernels whereas the large weight matrix in a fully
connected layer corresponds to large common data. The approach used in
NeuroTrainer is to buffer copies of small common data across PEs
and stream partitions of large data (e.g. inputs to a layer) from the local vault. Alternatively, with respect to large common weight matrices, they can be broadcast from a shared vault to all PEs, while partial weight matrices are stored across PEs. These two classes of data flows are illustrated in Fig.~\ref{fig:hybrid_data_flow}.

%  shows different data flow schemes. For small common input, 

% \textbf{Small common data.}
% Small common data can be stored in the buffer of PE and reused several times. For example, in convolution feedforward, kernels (W) are pre-stored in PE's buffer and used multiple times to cover all inputs. Non-shared data (Ex: X) needs to be fed from vault to PE directly (Fig.~\ref{fig:hybrid_data_flow} (a)). This data flow is valid when the amount of common data can be fitted into buffer of PE. For example, if a single kernel is bigger than buffer capacity (convolution weight update), this data flow is not valid.

% \textbf{Large common data.}
% If the amount of common data is large, pre-storing them in buffer is not feasible. Therefore, each PE needs to receive two inputs from vaults. In~\cite{kim2016neurocube}, all vaults push partial common data to all PEs simultaneously and it causes huge traffic congestion in NoC and low throughput for fully connected layer. To avoid the congestion, separating data paths for \textit{common data} and \textit{non-common data} is considered. A single vault stores common data and broadcasts to all PEs while each PE receive non-common data from dedicated vault. For example, in fully connected layer feedforward, input (X) is stored in a single vault and broadcast to all PEs and partial weight matrix (W) is directly delivered from vault to connected PE (Fig.~\ref{fig:hybrid_data_flow} (b)).

% \textbf{Data rearrange.}
The \textit{data rearranging} among vaults is required to dynamically change data flow from one type of layer to another. However, in a DNN, a set of convolution layers is followed by a set of fully connected layers; therefore rearrange is required only once in both feedforward and backpropagation. 

%If the computation result is \textit{common data} for next layer, all results from all PEs should be collected in a single vault. Although sharing a single data path by all PEs can be bottleneck of the system, as the amount of output is small the overall latency is not significantly degraded.  

\subsection{Programmable Memory Address Generator (PMAG)}\label{subsec:PMAG_arch}

\begin{figure}[t]
  \centering
  \includegraphics[width=1\columnwidth]{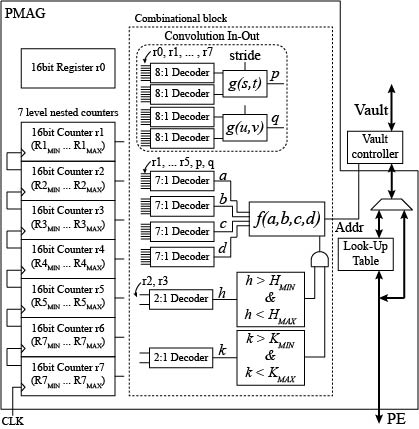}
  \caption{Block diagram of PMAG} 
  \label{fig:PMAG}
\end{figure}

The programmable memory address generator (PMAG) controls the data flow by providing memory address to the vault controller for read and write, and pushing the data through the~\proposedarch. The PMAG is composed of 7-level nested counters ($r1 ... r7$), combinational logic to generate address, and decoders to assign counter values as input of combinational logic (Fig.~\ref{fig:PMAG}). The PMAG also computes the non-linear function (and its derivative) by using look up tables [LUTs, for $f(x)$ and $f'(x)$] for (a) activation function (ReLu, tanh, etc.) or (b) exponential/logarithm for softmax and cross-entropy layer.

\begin{figure}[t]
  \centering
  \includegraphics[width=1\columnwidth]{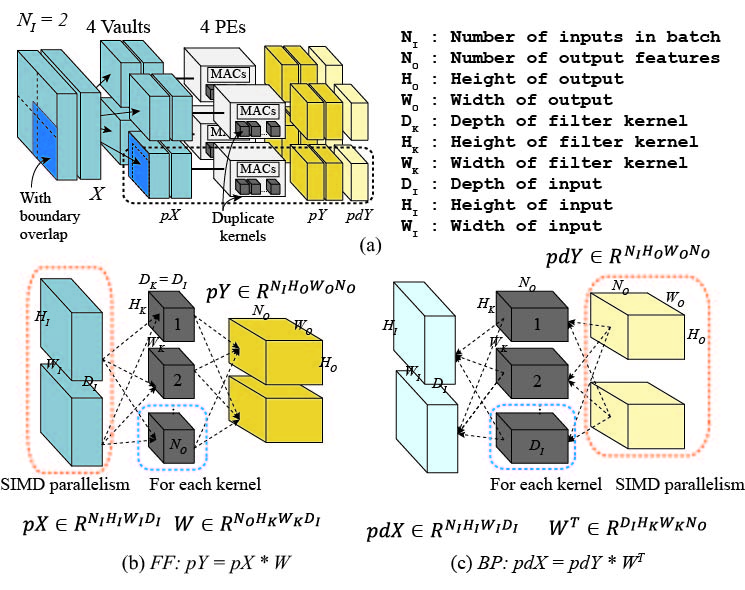}
  
    \caption{Convolution ($X$ and $W$). (a) X is partitioned into 4 $pX$ for 4 PEs, (b) Convolution feedforward, (c) Convolution backpropagation.} %Number of samples in batch ($N_I$) is 2} 
    
  \label{fig:conv_ff_bp}
\end{figure}

\textbf{Convolution Feedforward / Backpropagation.} 
Fig.~\ref{fig:conv_ff_bp} shows input $X$ is partitioned into 4 $pX$ with boundary overlap for convolution (assume 4 PEs). As kernel size is small, kernels are duplicated into all PE's buffers. For each kernel (outer most loop is $N_O$), $N_{MAC}$ inputs are processed in parallel (SIMD). For backpropagation, transpose of $W$ ($W^T$) is required and it can be handled in PE without reshaping data in the buffer of PE. It will be explained in Section~\ref{subsec:PE_op}.

\begin{figure}[t]
  \centering
  \includegraphics[width=1\columnwidth]{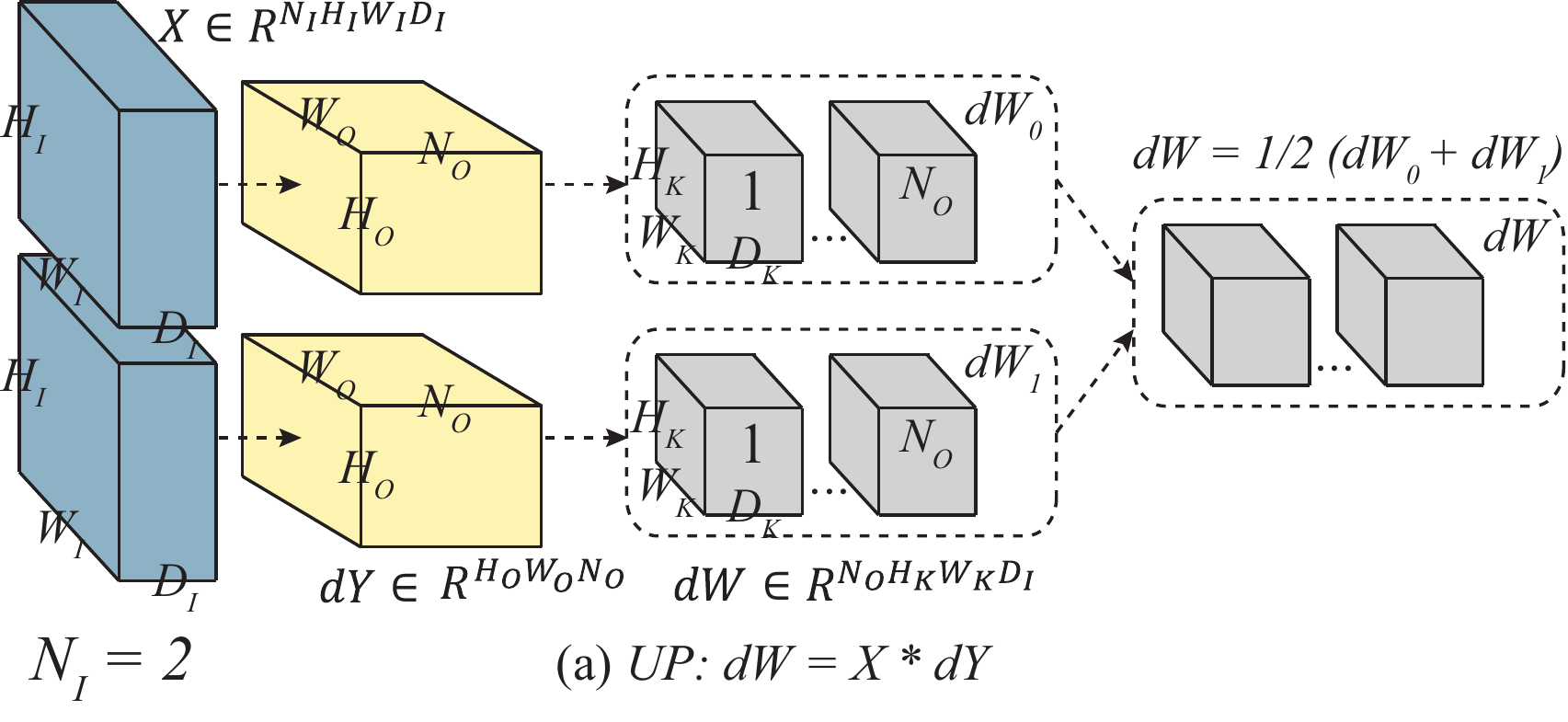}
  \caption{Convolution weight update when $N_I$ is 2. It will generate $dW_0$ and $dW_1$ by $dW_i = X_i * dY_i$. Final $dW$ is average of $dW_0$ and $dW_1$.} 
  \label{fig:conv_up}
\end{figure}

\textbf{Convolution Weight-update.} 
After generating $dY$, $dW$ is needed to update weights. Fig.~\ref{fig:conv_up} shows convolution weight update when $N_I$ is 2. For each sample ($X_i$), $dW_i = X_i * dY_i$ is computed, and final $dW$ is computed by averaging all $dW_i$s. Although weight update is also convolution between $X$ and $dY$, the kernel size ($W_O$ by $H_O$) is similar to the input size ($W_I$ by $H_I$). Due to large kernel ($dY$), partitioning input ($X$) with boundary overlap (Fig.~\ref{fig:conv_ff_bp}) is inefficient and duplicating $dY$ into all PEs is impractical. Therefore, we convert convolution with large kernel to matrix matrix multiplication by lowering convolution similar to how cuDNN performs convolution~\cite{chetlur2014cudnn} (Fig.~\ref{fig:conv_up} (b)). Although drawback of lowering is increasing memory requirement from $X_i$ to $X_{Mi}$, in-memory computation in~\proposedarch~can resolve the memory challenge.

\begin{figure}[t]
  \centering
  \includegraphics[width=0.8\columnwidth]{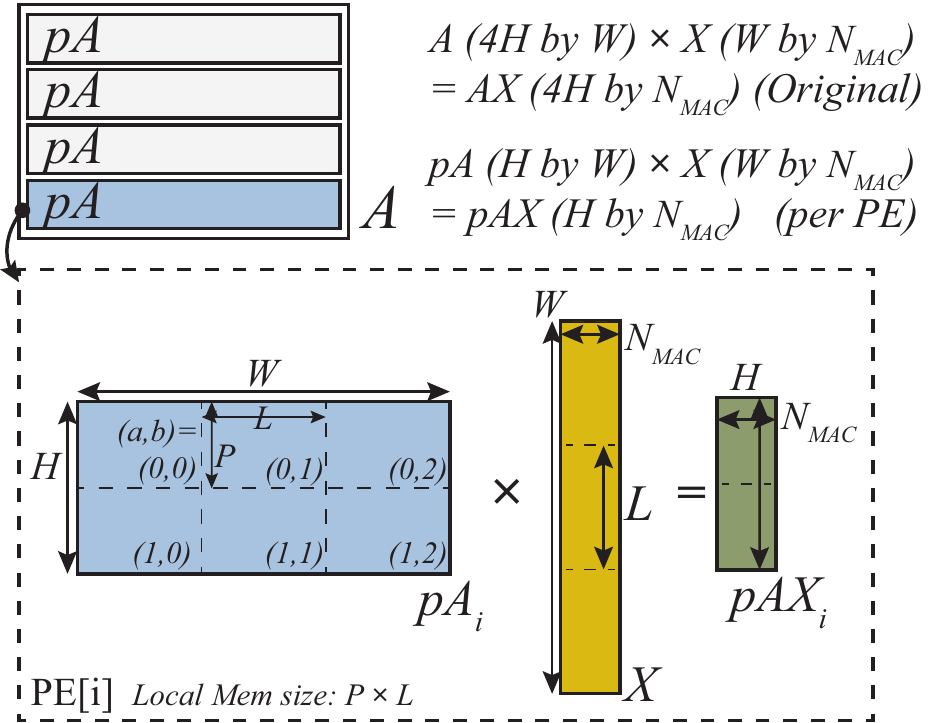}
  \caption{Matrix-matrix multiplication using 4 PEs. Each PE computes $pA \times X = pAX$.} 
  \label{fig:mat_mat_mul}
\end{figure}

\textbf{Matrix-matrix multiplication.} 
The main operation of fully connected layer or recurrent layer is matrix-matrix multiplication ($A \times X = AX$)~\cite{hochreiter1997long,chung2014empirical}. Fig.~\ref{fig:mat_mat_mul} shows that $A$ is divided into 4 $pA_{i}$ ($i$: PE index) row-wise and how a single $pA_{i}$ is partitioned to small blocks (each size is $L \times P$), which is fitted into half size of buffer in PE (double buffering). As explained in Section~\ref{subsec:hybrid_data_flow}, two data paths operate in matrix matrix multiplication (large common data) and the PMAG with common data vault and the PMAG with independent data vault are programmed separately. Fig.~\ref{fig:mat_mat_mul} shows that $pA_i$ is partitioned into 3 by 2 blocks. After processing first 3 blocks of $pA_{i}$ and $X$, a block of $pAX_{i}$ is generated (size = $N_{MAC}$ by $H$). The $pAX_{i}$ needs to be delivered to common data vault.

\begin{figure}[t]
  \centering
  \includegraphics[width=1\columnwidth]{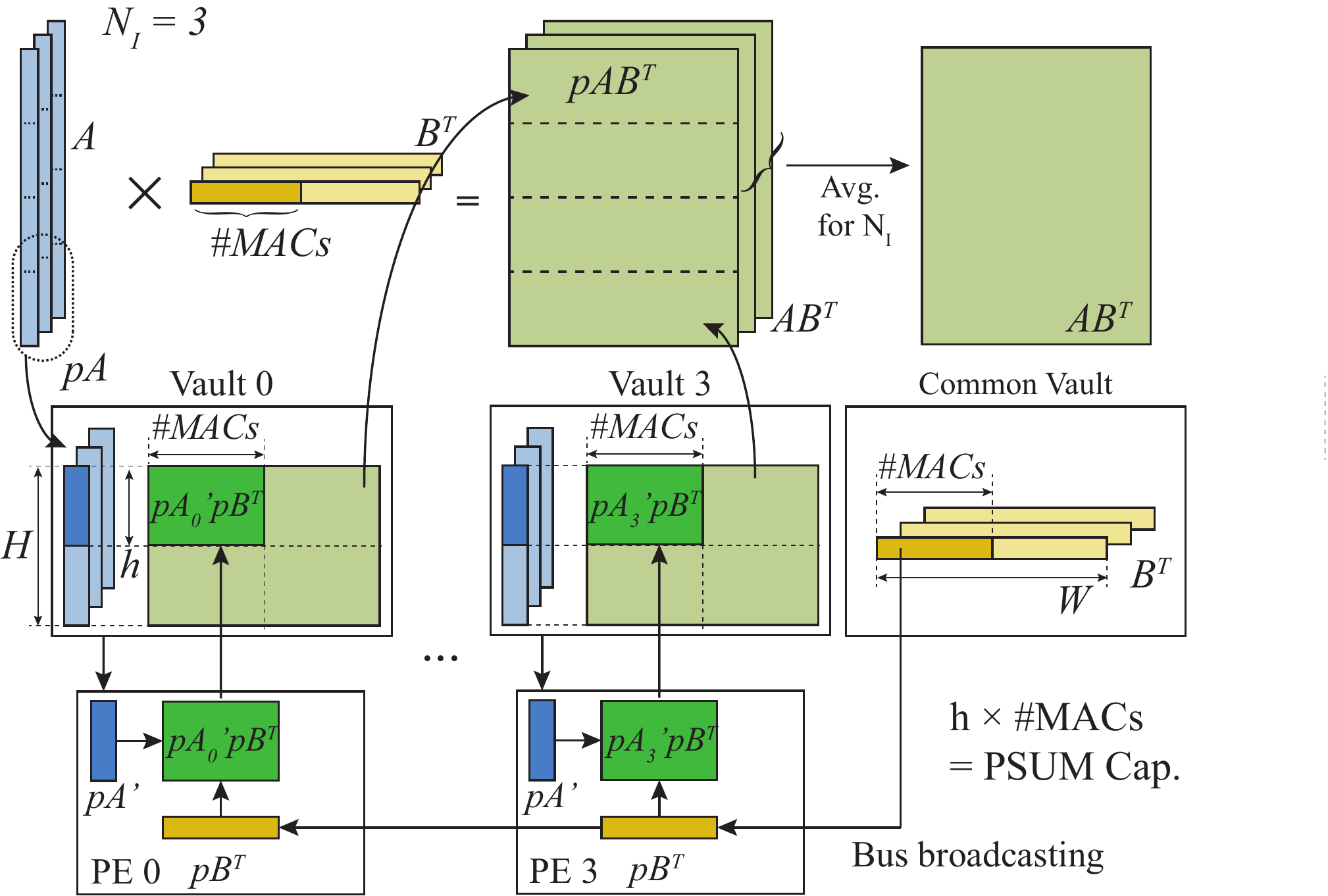}
  \caption{Vector-vector outer multiplication using 4 PEs. Each PE computes $pA \times B^T = pAB^T$.} 
  \label{fig:vec_vec_outer}
\end{figure}

\textbf{Vector-Vector Outer product.} 
For weight update in FC layer, for each sample in batch, input ($X$) and gradient ($dY$) need to be multiplied to generate $dW$. Contrast to matrix-matrix multiplication, $N_i$ samples cannot be unlooped in SIMD level. In other words, this operation should be repeated $N_i$ times and $dW$ needs to be averaged. Fig.~\ref{fig:vec_vec_outer} shows vector-vector outer product using 4 PEs. Vector $A$ is divided into 4 vaults ($pA_i$, size = $H$) and $B$ will be stored in common data vault and will broadcast. The operation inside PE is similar to that of matrix matrix multiplication, however, the output ($pA'pB^T$) does not need to be merged to common vault since it's gradient of weight in FC layer; therefore it's written back to dedicated vault.

\textbf{Data Preparation} 
Fig.~\ref{fig:conv_ff_bp} (a) shows that convolution with 4 PEs generates 4 $pY$s in parallel. If it is the last convolution layer before fully connected layer, the outputs of convolution layer should be \textbf{merged} into common data vault before to be broadcast in matrix matrix multiplication (Fig.~\ref{fig:hybrid_data_flow}). The order of PE to send data is pre-determined in the BUS. Based on this order, PMAG connected to common data vault also knows the portion in the merged data ($PW$, $PH$). In similar way, data from common data vault is also \textbf{partitioned} to all other vaults. 

\textbf{Add/remove zero boundary.}Before convolution, input needs to be \textbf{zero padded} on the boundary to return same sized output based on the kernel radius $r$.

\subsection{Processing Elements (PE)}
\begin{figure}[t]
  \centering
  \includegraphics[width=0.6\columnwidth]{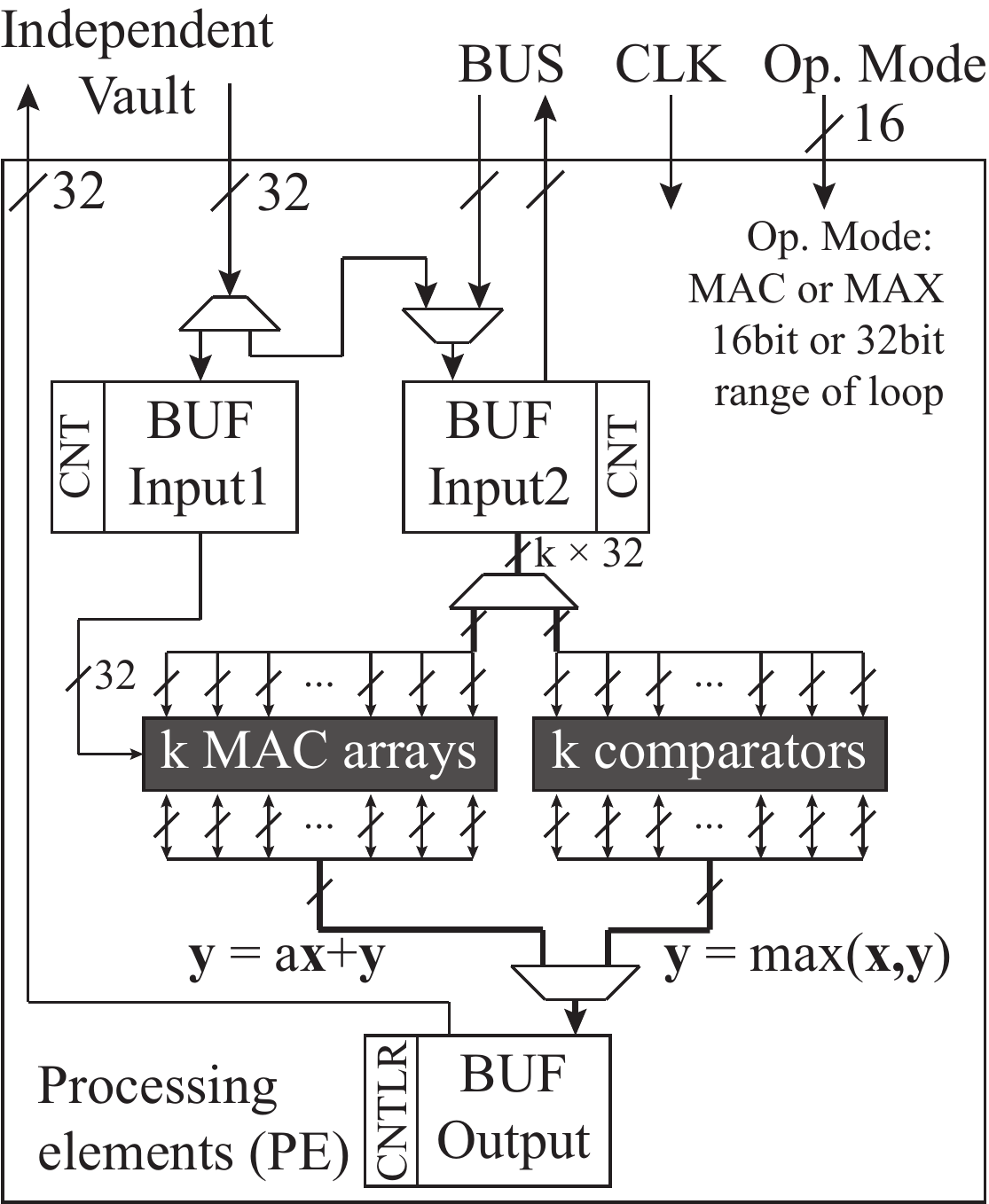}
  \caption{Block diagram of PE composed of three local buffers (input1, input2, and output), $k$ MAC, $k$ comparators.} 
  \label{fig:pe}
\end{figure}

%Fig.~\ref{fig:pe} illustrates PE[$i$] paired with Vault[$i$] and connected with data bus. 
A processing element is composed of a $k$ MACs array, $k$ comparators, and three local buffers: two input buffers, one output buffer (partial sum) (Fig ~\ref{fig:pe}). Similar to PMAG, PE also needs to be programmed before the main computation.

%A key innovation in the PE is to use MACs with variable precision (16bit or 32 bot)  fixed point arithmetic with stochastic rounding to achieve similar training accuracy as floating point design, but with higher efficiency. Similar to PMAG, PE needs to be programmed before main computation to select (i) MAC (most of multiplication and accumulation) or comparison (MAX pooling), (ii) the bit-precision modes - 16 bit for feedforward, 32bit for backpropagation and weight update, and (iii) range of loops for counters attached to three buffers as address generator.

%\begin{figure*}[t]
  %\centering
  %\includegraphics[width=0.8\textwidth]{./Memory_Mapping.eps}
  %\caption{Data mapping in DRAM to eliminate data movement between PEs} 
  %\label{fig:data_map}
%\end{figure*}

\subsubsection{Local Buffers} \label{subsubsec:cache}

To avoid stall of PE (idle mode) due to lack of operands (inputs), we use two inputs and output buffers in PE. 
%As explained in Section~\ref{subsec:hybrid_data_flow}, a single PE[$i$] has two data ports connected to memory channel separately. One is directly connected to independent vault [$i$] and one is connected to BUS network to communicate with common vault. 
%The amount of data stream to each buffer is defined as half of buffer size; therefore 
For each buffer, while half of memory is consumed by MACs (computing operation), rest of buffer can be filled simultaneously (double buffering). All local buffers have address generator based on nested counters. In Fig.~\ref{fig:pe}, CNT2 is two level nested 16bit counters and CNT1 is single level 16bit counter. 
Data stream between DRAM and PE ends with END-MARK (0xFFFF for 16bit case, 0xFFFFFFFF for 32bit case) 
% rather than handshaking signal between sender and receiver as PEs can have different latency from a single state vault in broadcasting mode. 
Computing in the PE starts only both input buffers are ready (half filled).

\subsubsection{Multiply and Accumulate Units}

\begin{table}[t]
\centering
\caption{Comparison of different fixed point MAC designs with IEEE 754 single precision floating point MAC. All designs are synthesized with 15nm FinFet~\cite{15nmlib} operating 2.5GHz.}
\label{table:MAC}
\begin{tabular}{|l|l|l|}
\hline
                    & Area ($um^2$)      & Power (mW)   \\ \hline
Float 32            & 2093.88         & 5.37         \\ \hline
Fixed 32/16         & 986.23 (-52\%)  & 2.27 (-57\%) \\ \hline
Fixed 32/16 SR~\cite{chainer2017improving}    & 2072.44 (+1\%)  & 5.79 (+7\%)  \\ \hline
Fixed 32/16 SR LO & 1578.71 (-24\%) & 3.78 (-30\%) \\ \hline
\end{tabular}

\end{table}

\begin{figure}[t]
  \centering
  \includegraphics[width=0.8\columnwidth]{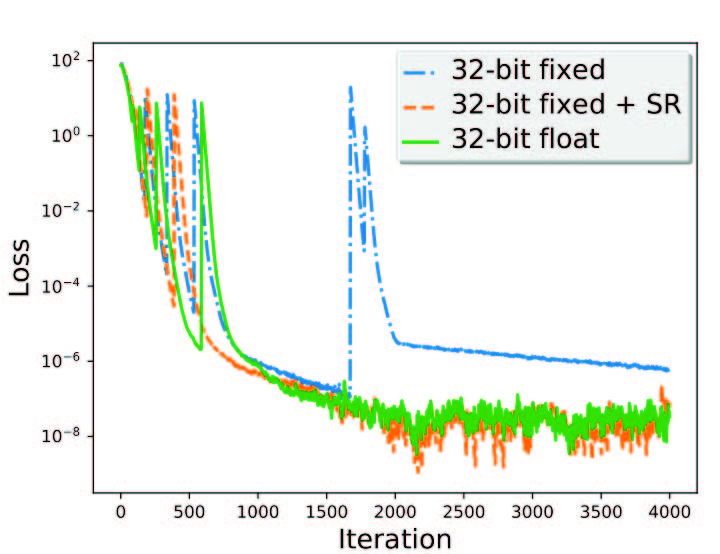}
  \caption{Training accuracy for RNN with different numeric representation (SR: Stochastic Rounding).} 
  \label{fig:training_acc}
\end{figure}

% Most of well known neural network layers' operations can be represented as a matrix vector multiplication since the state of neuron is computed as below:
% $$ y_{i} = \sum_{j=1}^{m} w_{i,j} \cdot x_{j}, $$
% where $y_i$ is the state of current neuron $i$ and $x_j$ is the output of neuron $j$ in previous or current layer.
As primitive arithmetic operator, a row of $k$ multiplier and accumulator (MAC) units is placed in a PE. 
%As low bit precisionlthough, 8 or 16 bit fixed point precision is enough for inference of deep learning~\cite{reagen2016minerva,kim2016neurocube,chen2016eyeriss,li2016high}, 
Although, reduced precision (16 bit fixed-point) is acceptable for inference (forward propagation), even 32bit fixed point in backpropagation and parameter update may result in inaccurate training in deeper network, in particular, the recurrent networks ~\cite{na2017chip} as illustrated in Fig.~\ref{fig:training_acc}. We should note that there is no accuracy degradation between $SR$ and $SR~LO$.
%which shows the training loss for a RNN with different precision. 
The stochastic rounding (SR) can be applied to overcome \textit{quantization} error in fixed point ~\cite{gupta2015deep,na2017chip}. 

%Therefore, we design MACs with variable precision (16bit or 32 bit) fixed point arithmetic with stochastic rounding to achieve similar training accuracy as floating point design, but with higher efficiency.

% Fig.~\ref{fig:training_acc} shows cost function of network during the training RNN. 32bit fixed point with stochastic rounding shows almost similar training trends with the trends of floating point case, while 32bit fixed point shows huge fluctuation in cost function and cannot converge as the result of floating point. It shows that using 32bit fixed point is not enough to train the deep network.

% In terms of hardware, four different MAC units are designed: 32bit floating point, fixed 32/16bit precision, Fixed 32/16 with stochastic rounding (SR), and Fixed 32/16 with stochastic rounding (SR) low overhead version (Table~\ref{table:MAC}).

%\begin{figure}[t]
%  \centering
%  \includegraphics[width=0.8\columnwidth]{./simd_mac.eps}
%  \caption{Multi-precision MAC operating mode (a): 32bit mode and (b): 16bit mode.} 
%  \label{fig:simd_mac}
%\end{figure}

\begin{figure}[t]
  \centering
  \includegraphics[width=0.6\columnwidth]{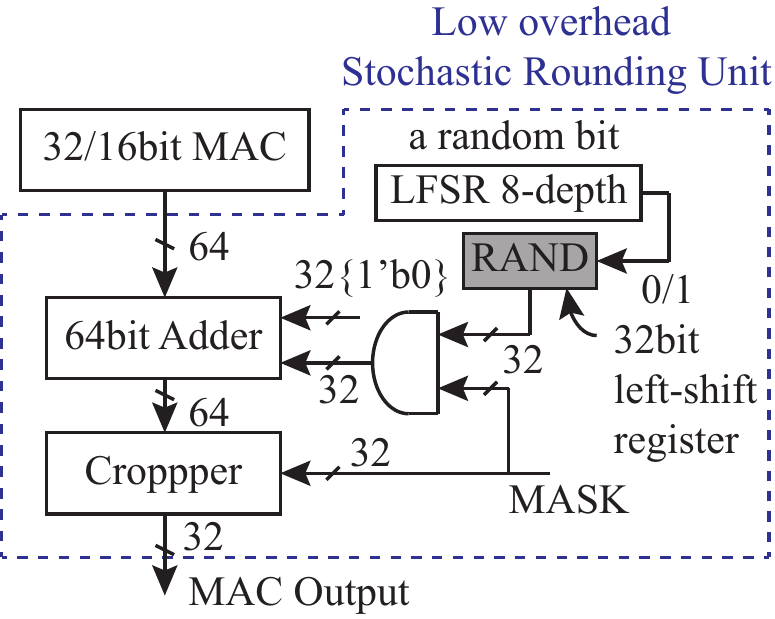}
  \caption{Fixed 32/16 + SR LO: Fixed 32/16bit MAC with low overhead stochastic rounding unit: a single LFSR and 32bit left shift register.} 
  \label{fig:mac_rand}
\end{figure}

Therefore, we design MAC to operate 1) two pairs of 16bit operands or 2) a pair of 32bit operands (Fixed 32/16). To add stochastic rounding, 64 random number generators are added~\cite{na2017chip} (Fixed 32/16 + SR). To reduce power/area overhead, we propose to add a single random number generator is used and it generates a single bit in every clock (\textbf{Fixed 32/16 + SR LO}, Fig.~\ref{fig:mac_rand})). Synthesis in 15nm FinFET~\cite{15nmlib} shows that proposed design provides higher energy-efficiency (Table~\ref{table:MAC}) while providing similar training accuracy as floating point design. The MAC operates in the 16bit mode without SR during inference (forward propagation). 

%The results show that the proposed design provides higher energy-efficiency than floating point design while maintaining training accuracy.

% First, MAC is designed to operate in two different precision modes: compute a pair of 32bit fixed point operands or two pairs of 16 bit fixed point operands in a single clock. Both 32bit and 16bit fixed point has 4 bit integer part and 28/12 bit for fractional part. Fig.~\ref{fig:simd_mac} shows operation multi-precision MAC operations. During the training, 16-bit state and weight are zero-padded and applied to the MAC. From 64-bit output, 32 bit output is cropped while maintain 4bit integer part. The result has 24bit fractional part (last 4 bits are zero). During the inference, two pairs of 16bit operands are delivered to MAC arrays. The result is two 32-bit outputs. It is cropped to two 16-bit outputs while 4-bit integer is maintained. (Fixed 32/16). 

% \paragraph{Fixed 32/16 + SR LP}  It feeds to LSB of 32bit left shift register RAND register (Fig.~\ref{fig:mac_rand}). After initial 32 cycles, RAND will deliver 32bit lenght random value in every cycle. By reducing the overhead of stochastic rounding unit, its power and area overhead is smaller than using floating point unit about 30\%. 

\begin{table*}[t]
\centering
\caption{Programming PMAG for Convolution and Fully connected layer}
\label{table:pmag_op}

\begin{tabular}{|c|c|c|c|c|c|c|c|c|c|c|c|c|c|c|c|}
\hline

\multirow{3}{*}{}   & \multicolumn{7}{c|}{\multirow{2}{*}{7 level nested counters}} & \multicolumn{4}{c|}{Conv. In - Out} & \multicolumn{4}{c|}{\multirow{2}{*}{f (a,b,c,d)}} \\ 
\cline{9-12}  & \multicolumn{7}{c|}{} & \multicolumn{2}{c|}{p} & \multicolumn{2}{c|}{q} & \multicolumn{4}{c|}{}  \\ 
\cline{2-16} & R1 & R2  & R3 & R4  & R5 & R6  & R7 & s & t & u & v & a & b & c & d \\ \hline

Conv-FF & $N_O$   & $H_O$     & $W_O$  & $N_I$  & $D_K$  & $H_K$  & $W_K$  & r2         & r6        & r3         & r7        & r4         & q          & p          & r5         \\ \hline
Conv-BP & $D_I$   & $H_I$     & $W_I$  & $N_I$   & $N_O$  & $H_K$  & $W_K$  & r2         & r6        & r3         & r7        & r4         & q          & p          & r5         \\ \hline
Conv-UP & 1    & $N_I$     & $H_O$  & $W_O$   & $D_I$  & $H_K$  & $W_K$  & r3         & r6        & r4         & r7        & q          & p          & r5         & r2         \\ \hline
\begin{tabular}[c]{@{}c@{}}FC-FF/BP\\ (C. Vault)\end{tabular} & H/P  & W/L    & P   & L    & K   & 1   & 1   & -          & -         & -          & -         & r4         & r2         & r5         & 0           \\ \hline
\begin{tabular}[c]{@{}c@{}}FC-FF/BP\\ (I. vault)\end{tabular}  & H/P  & W/L    & P   & L    & K   & 1   & 1   & -          & -         & -          & -         & r4         & r3         & r2         & r1          \\ \hline
\begin{tabular}[c]{@{}c@{}}FC-UP\\ (C vault)\end{tabular}    & H/h  & $\frac{W}{N_{MAC}}$ & $N_I$  & $N_{MAC}$ & 1   & 1   & 1   & -          & -         & -          & -         & r4         & r3         & r2         & r1          \\ \hline
\begin{tabular}[c]{@{}c@{}}FC-UP\\ (I. vault)\end{tabular}     & H/h  & $\frac{W}{N_{MAC}}$ & $N_I$  & h    & 1   & 1   & 1   & -          & -         & -          & -         & r4         & r3         & r2         & r1         \\ \hline

\end{tabular}
\begin{flushright}
	a, b, c, d, p, q, s, t, u, and v are labels used in Fig.~\ref{fig:PMAG}.\\
  R1 $\sim$ R7: maximum value of r1 $\sim$ r7 loop. (minimum value are all zero)
\end{flushright}
\end{table*}

% Please add the following required packages to your document preamble:
% \usepackage{multirow}
\begin{table*}[t]
\centering
\caption{Programming PMAG for data rearranging and data preparation}
\label{table:pmag_data}
\begin{tabular}{|c|c|c|c|c|c|c|c|c|c|c|c|c|c|c|c|}
\hline

\multirow{3}{*}{}  & \multicolumn{3}{c|}{\multirow{2}{*}{\begin{tabular}[c]{@{}c@{}}7 level\\ nested counters\end{tabular} }} & \multicolumn{4}{c|}{Conv. In - Out}  & \multicolumn{4}{c|}{\multirow{2}{*}{f (a,b,c,d)}} & \multicolumn{4}{c|}{\multirow{2}{*}{Two comparators}} \\ 

\cline{5-8} & \multicolumn{3}{c|}{}  & \multicolumn{2}{c|}{p} & \multicolumn{2}{c|}{q} & \multicolumn{4}{c|}{}  & \multicolumn{4}{c|}{} \\ 

\cline{2-16} & R1  & R2    & R3  & s & t & u & v & a & b & c & d & h & H & k & K               \\ \hline
Merge                                                                  & $D_I$   & $PH_I$    & $PW_I$    & -          & -         & -          & -         & r3         & r2         & r1         & 0          & -       & -               & -       & -               \\ \hline
Partition                                                              & $D_I$   & $H_I$     & $W_I$     & -          & -         & -          & -         & 0          & 0          & 0          & 1          & r2      & \begin{tabular}[c]{@{}c@{}}$ 0 \sim$ \\ $PH_I$\end{tabular}       & r3      & \begin{tabular}[c]{@{}c@{}}$ 0 \sim$ \\ $PW_I$\end{tabular}      \\ \hline
Add pad                                                                & $D_I$   & $PH_I$    & $PW_I$    & r3         & r         & r2         & r         & p          & q          & r1         & 0          & r3      & \begin{tabular}[c]{@{}c@{}}$ r \sim$ \\ $r+W_I$\end{tabular}   & r2      & \begin{tabular}[c]{@{}c@{}}$ r \sim$ \\ $r+H_I$\end{tabular}     \\ \hline

\begin{tabular}[c]{@{}c@{}}Remove\\ pad \end{tabular} & $D_I$   & $PH_I$    & $PW_I$    & -          & -         & -          & -         & r3         & r2         & r1         & 0          & r3      & \begin{tabular}[c]{@{}c@{}}$ r \sim$ \\ $r+W_I$\end{tabular}     & r2      & \begin{tabular}[c]{@{}c@{}}$ r \sim$ \\ $r+H_I$\end{tabular} \\ \hline
\end{tabular}
\begin{flushright}
  a, b, c, d, p, q, s, t, u, and v are labels used in Fig.~\ref{fig:PMAG}.\\
	R1 $\sim$ R3: maximum value of r1 $\sim$ r3 loop. (minimum value are all zero) \\
  R4 $\sim$ R7: 1

\end{flushright}

\end{table*}

\begin{table}[t]
\centering
\caption{PE Program for computing operations}
\label{table:pe_program}
\begin{tabular}{|c|c|c|c|}
\hline
        & Bit & CNT2   & CNT1     \\ \hline
Conv-FF & 16  & $H_K, W_K$ & $W_K \times H_K$ \\ \hline
Conv-BP & 32  & $H_K, W_K$ & $W_K \times H_K$ \\ \hline
Conv-UP & 32  & $P, L$   & $L$            \\ \hline
FC-FF   & 16  & $P, L$   & $L$            \\ \hline
FC-BP   & 32  & $P, L$   & $L$            \\ \hline
FC-UP   & 32  & $h$      & $1$            \\ \hline
\end{tabular}
\begin{flushright}
  $H_K, W_K$: dimension of convolution kernels \\
  $P, L$: dimension of partial matrix (Fig.~\ref{fig:mat_mat_mul})\\
  $h$: length of partial vector (Fig.~\ref{fig:vec_vec_outer})

\end{flushright}
\end{table}

\subsubsection{Comparator Unit}
Since MAX operation is required only for the max-pooling inference, 16 bit fixed point comparators are placed in PE. Based on pooling radius ($r$), $r^2$ data are streamed into controller, and the comparator unit returns the maximum value and its ID for backpropagation.

\subsubsection{PE Operation} \label{subsec:PE_op}

After two input buffers are filled (BUF Input 1 and BUF Input 2), BUF Input 1 pushes one 32bit input (one 32 bit operand or two 16 bit operands) while BUF Input 2 pushes $k$ ($N_{MAC}$) 32bit inputs. For MAC operation (all cases except max pooling), $k$ MAC arrays compute $\textbf{y} = a\textbf{x} + \textbf{y}$, where \textbf{x} and \textbf{y} are vectors, which length is $k$ (32bit) or $2k$ (16bit). For MAX operation (max pooling), $k$ comparator returns max value as $\textbf{y} = max(\textbf{x},\textbf{y})$.

\textbf{Convolution.}
In convolution, $k$ inputs are processed by $k$ MACs in parallel (SIMD level). Therefore, kernels are stored in BUF Input 1 and $k$ inputs are stored in BUF Input 2. If $k$ inputs cannot be stored in BUF Input 2 due to capacity issue, $k$ subsets of $k$ inputs are stored and newly required input ($k \times H_k$) is updated during the operation similar to~\cite{chen2016eyeriss}. For convolution backpropagation, $W^T$ is easily obtained by sweeping counter values in CNT2 attached to BUF Input 1.

\textbf{Matrix-Matrix multiplication.}
Similar to convolution, $k$ inputs are processed in parallel (SIMD level). Therefore, partial weight matrix is loaded in BUF Input 1 and $k$ partial inputs are stored in BUF Input 2. After consuming one partial weight matrix [a,b] ($P$ by $L$ in Fig.~\ref{fig:mat_mat_mul}), next partial weight matrix [a,b+1] is processed. Similar to convolution, $W^T$ is obtained by sweeping counter values in CNT2 attached to BUF Input 1.

\textbf{Vector-Vector outer product.}
In fully connected update, $k$ inputs cannot be processed in parallel. In computing $AB^T$, $A$ is loaded in BUF Input 1 and $B$ is loaded in BUF Input 2. In other words, $k$ elements of $B$ is delivered into $k$ MAC units in a single clock (Fig.~\ref{fig:vec_vec_outer}).

\subsection{BUS Interface} \label{sec:bus}

%\begin{figure}[t]
%	\centering
%	\includegraphics[width=\columnwidth]{./bus_diagram.eps}
%	\caption{Block diagram of bus interface between state vault and all PEs.} 
%	\label{fig:bus}
%\end{figure}

Bus interface has two operation modes controlled by common data vault:
broadcasting to all PEs and merging data into common data vault from all PEs. 
The BUS and PE communicates using three-way handshaking (REQ-ACK-SEND) for both operations. Broadcasting mode is set when all PE can take data (input buffer is ready) and during the broadcasting, any REQ from PE is ignored (broadcasting is prior to merging mode). During the merging mode, all PE send REQ and get ACK from the bus based on predetermined priorities among PEs. Although all 15 PEs request BUS for writing-back simultaneously, the impact of latency of entire writing back (for 15 PEs) on the throughput can be minimized as PE's computing latency dominates entire computing latency. The bus architecture is designed and synthesized to guarantee a bandwidth same as that of a single vault (10GBps). We use 4 stage pipe-lined BUS interface~\cite{opencore}; it takes 4 clock cycle between a vault to any PE.
%Although more complex (2D mesh network of routers) and faster interface can be considered, the bandwidth of a single vault will become bottleneck in the system. Therefore, 10GBps BUS is good enough for~\proposedarch. 

\section{Programming}\label{programing}
\begin{figure}[t]
	\centering
	\includegraphics[width=0.8\columnwidth]{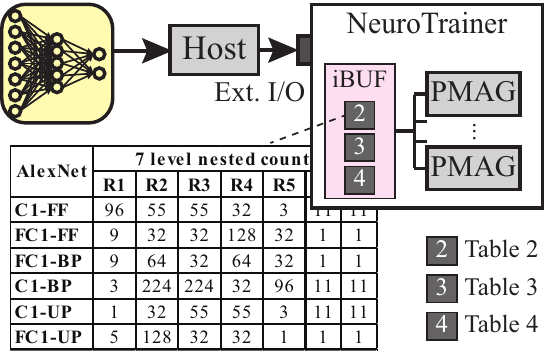}
	\caption{Programming NeuroTrainer by host for given DNN.} 
	\label{fig:programming_flow}
\end{figure}
 Following the discussions in Section ~\ref{subsec:PMAG_arch}, Table~\ref{table:pmag_op} and Table~\ref{table:pmag_data} summarize the PMAG programming which includes setting range of 7 nested loops ($r1 \sim r7$) and connecting counter values to combination logic for different operations. For matrix-matrix multiplication (FC-FF/BP) and vector outer product (FC-UP), C.Vault is the programming value for PMAG attached to common data vault and I.Vault is the programming value for PMAG attached to independent data vault. Similar to PMAG, PE needs to be programmed to set: 1) operation type: MAC or MAX, 2) bit precision mode for MAC: 16 bit or 32 bit with/without SR , and 3) loop range for address generator for local buffers. Based on the discussions in Section ~\ref{subsec:PE_op}, Table ~\ref{table:pe_program} summarizes the inputs for PE programming for different operations. In essence, the preceding three tables defines the instruction set architecture of the~\proposedarch.

Given a DNN, the host first generates the preceding three tables. 
%Once the instruction tables for a given DNN is generated, host can perform layer-by-layer programming, similar to ~\cite{kim2016neurocube}, where tables values for each layer are loaded into the PMAG and PEs. Once the layer is computed, the values for the next layer is programmed. 
Fig. \ref{fig:programming_flow} illustrates the programming process of the~\proposedarch. To enable \textit{autonomous operation} of the~\proposedarch, we embed an on-chip instruction buffer (iBuffer) to store the preceding three tables (Figure ~\ref{fig:training_efficiency}(a)). Given a DNN, the host generates the preceding three tables and loads them in the iBuffer. During execution the layer-wise operation is controlled by the iBuffer (using a layer counter). To estimate the size of the iBuffer, consider that for a network with $N$ layers, we need to program~\proposedarch~$\sim 4N$ times (Feedforward, backpropagation, weight update, data preparation if needed). Each time the amount of data for programming is 22Byte (18Byte for PMAG and 4Byte for PE). Therefore, a 16KB memory is sufficient as iBuffer and it can cover 186 layers. The latency of programming the iBuffer through HMC external interface is negligible compared to loading the input data.
 % As an example, Table~\ref{table:alexnet_program_ex} shows actual parameters need to be programmed into PMAG for C1, C2, FC1, and FC3 layer in Alexnet~\cite{krizhevsky2012imagenet} for feedforward, backpropagation, and weight update. In Alexnet feedforward, between C5 and FC1 layer, data needs to be merged (FC1 Prep) and data needs to be partitioned (C5 Prep) during the backpropagation. Once the iBuffer is loaded the cycle-level simulation is performed. 

\section{Simulation Results} \label{simul}

\subsection{Performance Analysis} \label{subsec:performance_simulation}

%% SAIBAL -- Please add few lines on how performance is simulated. 
The performance of the~\proposedarch~is simulated using cycle-level simulator. All simulation results is based on minibatch size 32, which is recommended minimum size of minibatch~\cite{BAIDU}. All MACs, comparators, buffer in PE, BUS interface, PMAG are synthesized operate at 2.5GHz to maximize the single vault's bandwidth.

\begin{figure*}[t]
  \centering
  \includegraphics[width=0.9\textwidth]{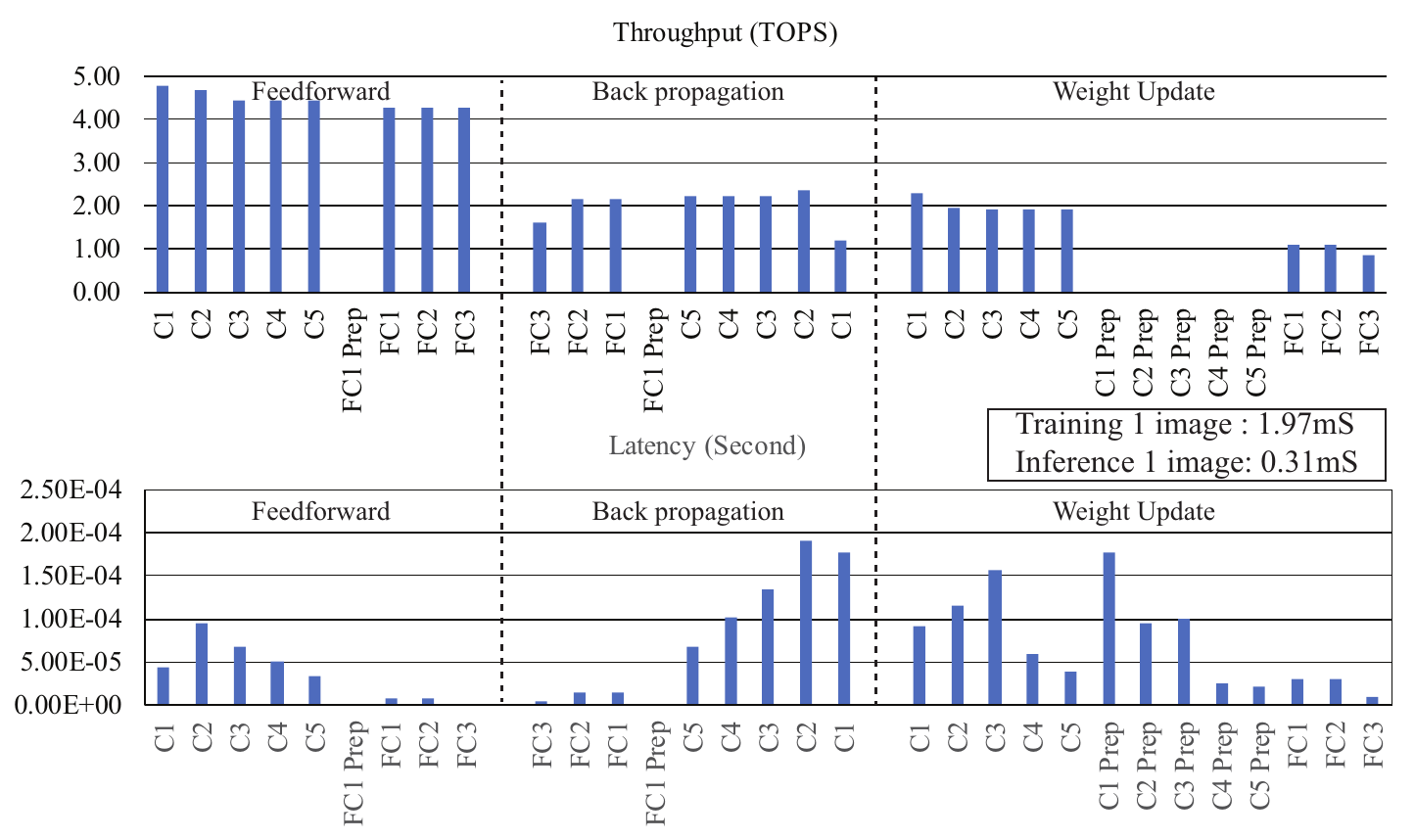}
  \caption{Simulation result for Alexnet in terms of latency (second) and throughput (Tera-Ops/sec: TOPS/s). C1 - C5: convolution layer, FC1 - FC3: fully connected layer, Prep: data preparation} 
  \label{fig:alexnet}
\end{figure*}

Fig.~\ref{fig:alexnet} shows throughput (TOPS/s) and latency (second) for a single input of each layer in Alexnet. 
%For layer-wise analysis, Alexnet~\cite{krizhevsky2012imagenet} is chosen since it has both convolution and dense connections and the number of layers are small enough (8) to analyze. 
For the one input image, inference took 0.31mS (3,228 images per second) and training took 1.97mS (507 images per second).

In feedforward phase, all convolution or fully connected layers shows similar throughput above 4.0 TOPS/s (4.2TOPS/s $\sim$ 4.7TOPS/s) which is close to the theoretical maximum for 16bit operation of our MAC (2.5GHz $\times$ 15 PEs $\times$ 32 MACs $\times$ 2 pairs inputs $\times$ 2 (multiplication and addition) = 4.8TOPS/s). %In other words, hybrid data flow is efficient both convolution with small kernel case and fully connected case.

For backpropagation and weight update, 32bit with stochastic rounding is used. Theoretical maximum throughput can be computed as 2.4 TOPS/s in the same manner. In backpropagation, FC3 (1.61 TOPS/s) and C1 layer (1.19 TOPS/s) show lower throughput than others. For FC3 backpropagation, the size of $dY$ is not large enough to hide latency of \textit{writing back} from all PEs to a single vault. In other words, the latency to generate output by iterating $dY$ times is shorter than writing back the output to common data vault; writing back becomes bottleneck in the system. For C1 layer, input dimension is $55 \times 55 \times 96$ and kernel dimension is $11 \times 11 \times 3$. It can be processed as convolution since kernel size is small enough to fit in the local memory; but efficiently due to large kernel size compared to input. %% SAIBAL - dont underdtand the last sentence

In weight update, C1 $\sim$ C5 shows about 1.98 TOPS/s by translating convolution as matrix multiplication in large kernel case. However, FC layer (vector vector outer multiplication) shows about 1.02 TOPS/s, which is the worst case due to high network traffic between PE and independent vault since there is no data re-usage. 

\begin{figure}[t]
  \centering
  \includegraphics[width=0.8\columnwidth]{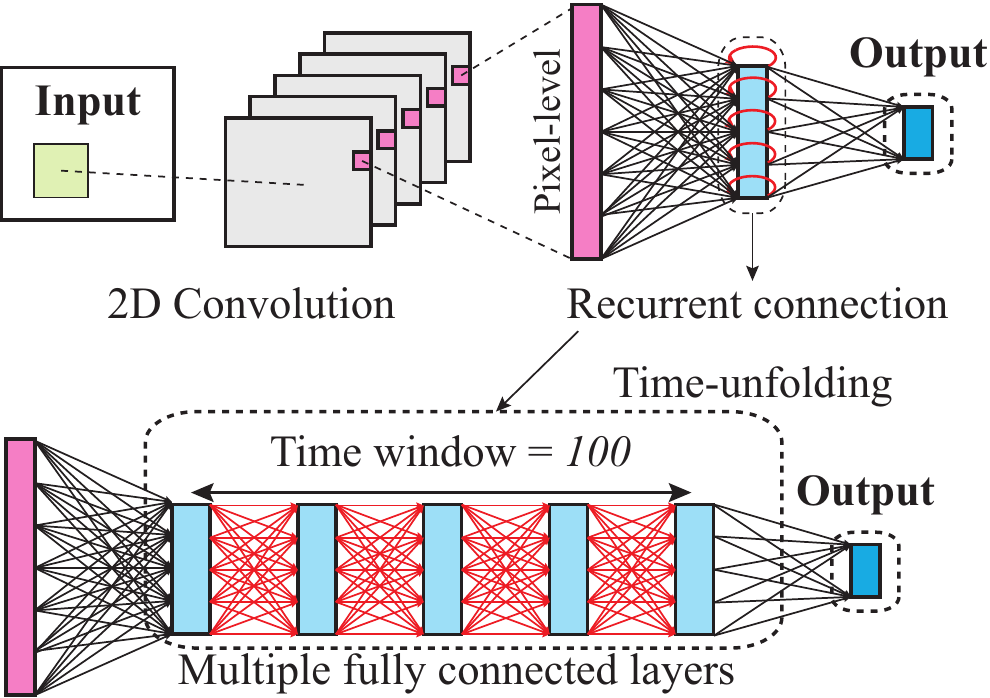}
%  \caption{Simplified example of DNN for generating sentences for image description~\cite{karpathy2015deep}. 2D convolutional layers extract features, fully connected layers classify the detected objects, and the recurrent layer generates the sequence of words. Recurrent connections (red lines) can be transformed into multiple
%fully connections after time unfolding.} 
  \caption{Simplified example of DNN for generating sentences for image description~\cite{karpathy2015deep}.} 
  \label{fig:dnn}
\end{figure}

\begin{figure*}[t]
  \centering
  \includegraphics[width=1\textwidth]{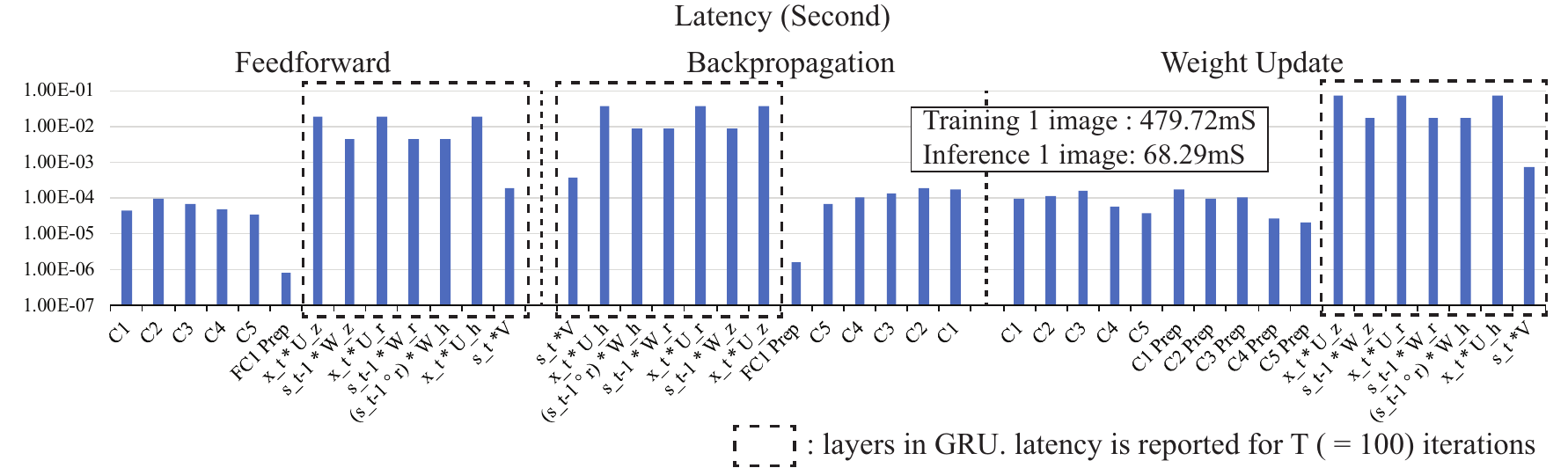}
  \caption{Latency analysis of each layer in DNN~\cite{karpathy2015deep}.} 
  \label{fig:dnn_simul}
\end{figure*}

To see the performance of more complex and deeper network, we evalaute a DNN for generating image description~\cite{karpathy2015deep}, image feature extraction part is implemented as Alexnet and RNN (GRU) is attached after $5^{th}$ convolution layer (Fig.~\ref{fig:dnn}).  A single GRU is composed of six fully connected layers for hidden neurons and one fully connected layer for output neurons. The number of input neurons in GRU is 43,264 and the number of hidden neurons in GRU is 10,000. We assume $T$ for DNN is 100. Fig.~\ref{fig:dnn_simul} shows latency of each layer in DNN explained earlier. For the recurrent layers (in the dashed box), the latency is computed considering time windows (latency to across all time unfolded $T$ layers); that's why it shows high latency than other layers. 
%The result emphasizes that achieving high throughput in fully connected layer is critical in temporal applications.

%%% SAIBAL - Please present some analysis from the figure as in the previous case. 

%In language processing with GRU, word embedding size (input neurons in GRU) is set as 3,000 and number of hidden neurons is set as 2,000~\cite{yandex}. For different time window ($T$), the length of time-unfolded GRU is determined. 

% Fig.~\ref{fig:gru} shows the latency to process a single image in GRU for both inference and training with different time window size ($T$). 

% A single GRU is composed of six fully connected layers for hidden neurons and one fully connected layer for output neurons. Therefore, when $T$ is 10, there are 60 fully connected layers for across 10 layers of hidden neurons and one last layer for output neurons. As time window increases, latency for both inference and training increases but throughput is independent with time window. Although the number of layers increases, the same size fully connected layers (number of connections) are repeated; therefore throughput is independent with the time window.

% In GRU, training latency is about 6.7 times of inference latency, which is reasonable since the number of operation for training is about 3 times of that inference approximately and backpropagation and weight update is 2 times slower than feedforward due to 32bit mode.

\begin{figure}[t]
  \centering
  \includegraphics[width=1\columnwidth]{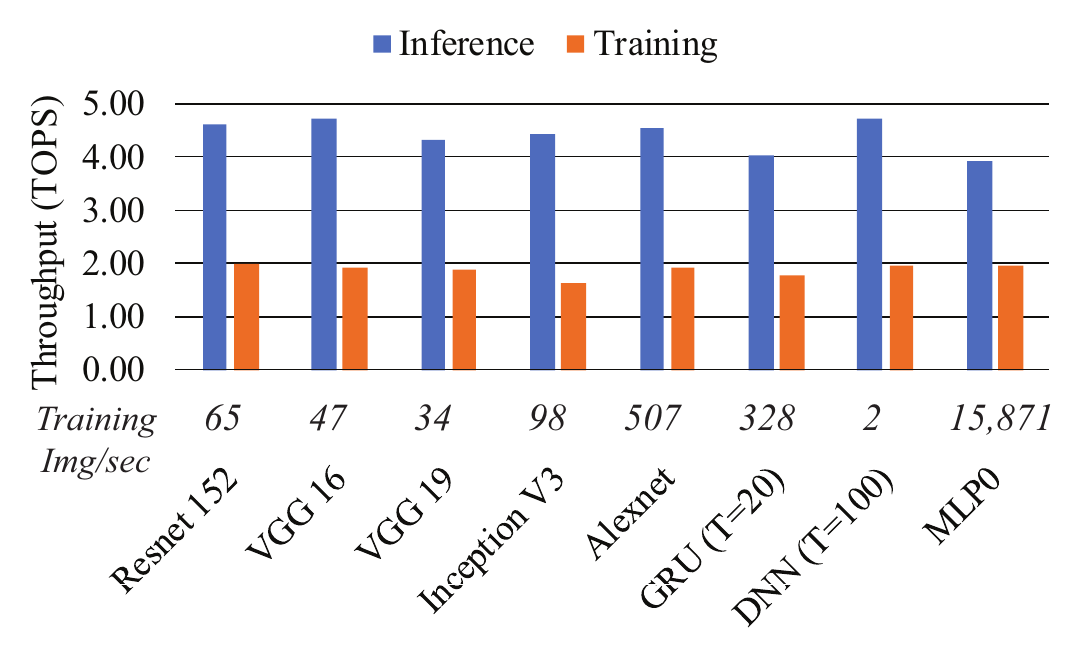}
  \caption{Simulation results for different benchmarks. } 

  \label{fig:summary}
\end{figure}

 % Y-axis represents system performance (throughput: Tera-Ops/sec). The numbers below each bar represent the number of inputs can be processed in a second for training.
 
Fig.~\ref{fig:summary} shows the throughput for various benchmarks including Resnet 152~\cite{he2016deep}, VGG 16, VGG 19~\cite{simonyan2014very}, Inception V3~\cite{szegedy2016rethinking}, GRU~\cite{chung2014empirical}, DNN for image description~\cite{karpathy2015deep}, and MLP0~\cite{TPU} are also tested.  Y-axis represent the throughput (TOPS/s) and the number on the X-axis represent the number of inputs can be trained in a second for each benchmark. For all benchmarks, inference shows $4.0 \sim 4.7$ TOPS/s and training shows $1.9$ TOPS/s. Further, \proposedarch~shows stable throughput (standard deviations less than 6\% of average) for training with all benchmarks of varying complexity.

\subsection{Synthesis and Power Analysis} \label{subsec:hw}

The computation fabric of the~\proposedarch, including the PEs, bus interface, and PMAG with the vault controller is synthesized using 15nm FinFet~\cite{15nmlib}. %In PE design, three 16KB SRAMs are placed and its area power value is referenced from~\cite{shafaei2014fincacti}. 
As vault controller is a proprietary design, a 32bit SDRAM controller~\cite{opencore} is adopted as a reference vault controller. Table~\ref{table:hw} summarizes average power across 8 different benchmarks and area overhead of each module in the system. Total power consumption of logic layer is 2.64W and area overhead (including vault controller) is 1.17$mm^2$. Even scaled up to compare with previous result synthesized in 28nm CMOS~\cite{jeddeloh2012hybrid}, total area is less than 5\% of footprint of fabricated HMC (68$mm^2$). Average DRAM die power is computed during the simulation using 3.7pJ/bit from~\cite{jeddeloh2012hybrid} and actual DRAM access pattern. The power densities of the logic die (0.039W/$mm^2$) and DRAM dies (0.030W/$mm^2$) in~\proposedarch~is well within the acceptable power densities ($1.5W/mm^2$, ~\cite{huang2010interaction}, of 3D stacked systems. 

From DRAM power consumption (2.03W), average memory bandwidth can be computed as 68.5GByte/sec (2.03W/3.7pJ/bit), which is lower than total aggregated memory bandwidth of HMC (16 vaults $\times$ 10GByte/sec). With batch size of 32, weights are reused 32 times. DRAM utilization can be increased by, 1) more MACs per PE, but requires larger partial sum SRAM and less efficient for the small batch (minimum batch size for most of DLs is 32) and 2) more PEs, but it requires a larger network among PEs and vaults.

\textit{On average,~\proposedarch~consumes 4.64W and delivers 1.89 TFLOPS throughput and 406 GFLOPS/W of efficiency during training (32bit) while maintaining high training accuracy.}

For HMC 2.0~\cite{hmc2}, performance is estimated (Table~\ref{table:prevwork}). With 32 vaults, 31 PEs can be placed; therefore throughput and logic power increases about twice. However, power of DRAM dies is same since total memory access is constant. Therefore, it shows 39\% gain in efficiency.

\begin{table}[t]
\centering
\caption{Power and Area analysis of~\proposedarch~synthesized in 15nm FinFet~\cite{15nmlib}.}
\label{table:hw}
\begin{tabular}{|l|c|c|}
\hline
                     & \textbf{Area ($mm^2$)} & \textbf{Power ($W$)} \\ \hline
\textbf{PE}          & 6.96E-02            & 1.55E-01           \\ \hline
\textbf{PMAG}        & 2.00E-03            & 3.16E-03           \\ \hline
\textbf{Vault Ctrl.} & 7.73E-04            & 4.27E-03           \\ \hline
\textbf{32bit Bus}   & 8.96E-03            & 3.70E-02           \\ \hline
\textbf{16KB: I-BUF} & 5.51E-03            & 1.02E-02           \\ \hline
\multicolumn{3}{l}{}                                            \\ \hline
\textbf{Logic die}   & 1.17E+00            & 2.65E+00           \\ \hline
\textbf{4 DRAM dies} &                     & 2.03E+00           \\ \hline
\end{tabular}
\end{table}

\subsection{Scalability to Multiple NeuroTrainers} \label{multi}

\begin{figure}[t]
  \centering
  \includegraphics[width=1\columnwidth]{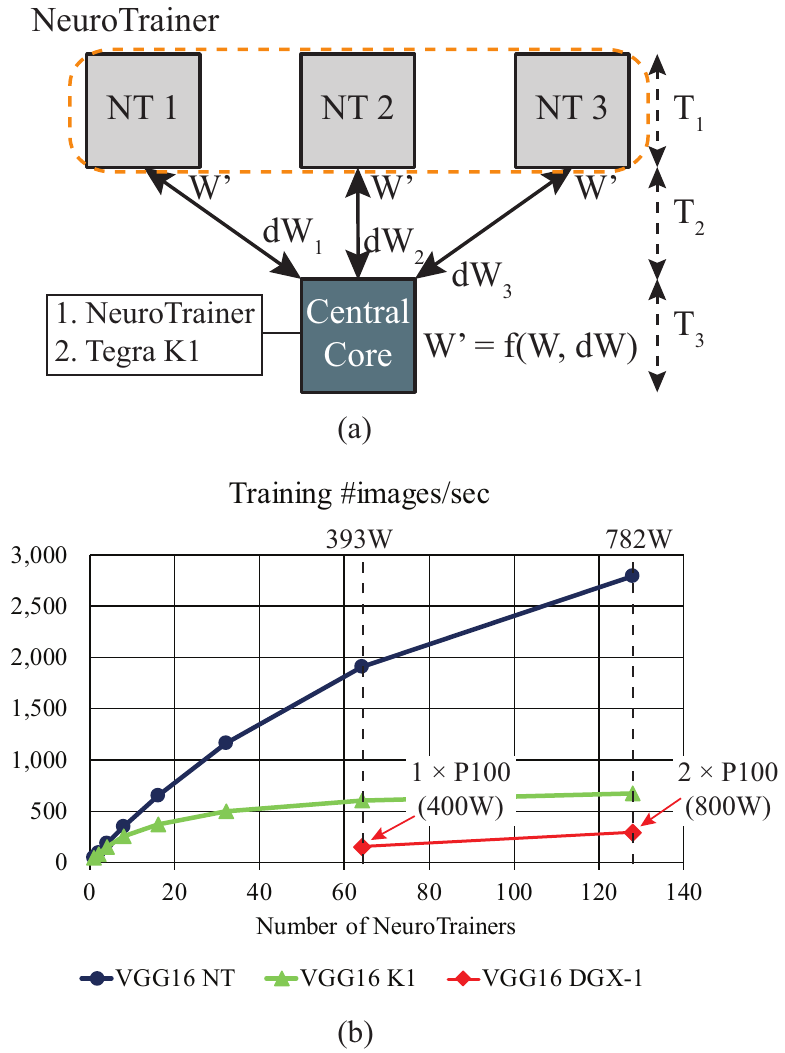}
  \caption{Scalability: (a) system of multiple~\proposedarch. (b) Performance for VGG16 with central core being~\proposedarch~(VGG16 NT), Tegra K1 (VGG16 K1), and DGX-1 (P100) (VGG16 DGX-1) .} 
  \label{fig:multi}
\end{figure}

The multiple ~\proposedarch~can be used in parallel for scalable training performance as illustrated in Fig.~\ref{fig:multi} (a).  As all~\proposedarch~take same latency ($T_1$) for training a minibatch, we propose to perform \textbf{synchronous} training~\cite{iandola2016firecaffe}. After training a single input batch, $N$~\proposedarch~delivers $dW$ to a central unit (latency = $T_2$). The central unit needs to take all $dW$ from $N$~\proposedarch~s, and generates new W ($W'$) following: $W' = \eta \times average(dW) + W $, where $\eta$ is learning rate. The above computation can be performed by another~\proposedarch. 
%However, there are more complex optimization algorithms to generate $W'$ such as AdaGrad~\cite{duchi2011adaptive} or Adam~\cite{kingma2014adam} and finding a better optimization algorithm is still on the research. 
However, to cover more generic approaches for weight update (e.g. AdaGrad~\cite{duchi2011adaptive} or Adam~\cite{kingma2014adam}), a software implementation, for example, using Tegra K1 (326 GFLOPS, 10W, 28nm)~\cite{nvidia2014k1} can also be considered.  

Fig.~\ref{fig:multi} (b) shows estimated training performance (number of images per second) of VGG 16~\cite{simonyan2014very} by different number of NeuroTrainers and two different types of central core.
% is studied: NT (\proposedarch) and K1 (NVIDIA Tegra K1~\cite{nvidia2014k1}). By $N$~\proposedarch s, $N \times 32$ images can be trained in $T_1$ (Batch size is 32). In $N \times T_2$, a central core takes all $dW$ and takes another $T_3$ to generate $W'$. $W'$ needs to be delivered to all~\proposedarch in $N \times T_2$. $T_2$ is estimated using HMC external bandwidth (160GByte/s).
% Thus, the number of trained images per second is: $NB/(T_1 + 2 N T_2 + T_3).$ 
Training performance using high-end GPU (NVIDIA DGX-1) is also reported. This estimated performance is computed based on peak FLOPS of each processing unit and HMC external BW. For example, Tegra K1 (326 GFLOPS) can update weights in 42.4mS for AlexNet (138M parameters) since it’s element-wise operation. The latency between a NeuroTrainer and K1 is 4.61mS (240 GByte/sec). If system is composed of 4 NeuroTrainers with a K1 as host, total latency is 63.1mS (training latency in NeuroTrainer) + 42.4mS $\times$ 4 (K1 needs update W using 4 dWs) + 2 $\times$ 4 $\times$ 4.61mS (round trip from a K1 to 4 NeuroTrainrs) = 269.58mS while training 4 $\times$ 32 samples. In the same manner, a single P100 in DGX-1 can train 150 images per second~\cite{benchmark} with 400W power consumption. For the same power budget, 64~\proposedarch~can operate in parallel and train 1,900 images delivering 13x speedup. The additional power consumption due to off-chip data movement estimated using HMC access energy of 10pJ/bit~\cite{jeddeloh2012hybrid}. Ultimately, the performance scaling in~\proposedarch~is limited by the off-chip latency showing need for better system architecture and faster off-chip network.

%The \proposedarch system can train more than 1,900 images (central core is~\proposedarch) showing ~\proposedarch~shows x13 performance than DGX-1 in same power budget. 

\section{Related Work} \label{related}

\begin{table}[]
\centering
\caption{Comparison with previous training accelerators}
\label{table:prevwork}
\begin{tabular}{|c|c|c|c|c|}
\hline
\textbf{Work} & \textbf{NC~\cite{kim2016neurocube}} & \textbf{NS~\cite{azarkhish2017neurostream}} & \textbf{SD~\cite{venkataramani2017scaledeep}} & \textbf{NT}  \\ \hline

\textbf{Bit}                                                    & 16 FI       & 32 FL       & \begin{tabular}[c]{@{}c@{}}16 FL\\     /32FL\end{tabular} & \begin{tabular}[c]{@{}c@{}}16 FI\\     /32 FI*\end{tabular}      \\ \hline
\textbf{\begin{tabular}[c]{@{}c@{}}Node\\ (nm)\end{tabular}}    & 15          & 28          & 14                                                        & 15                                                               \\ \hline
\textbf{\begin{tabular}[c]{@{}c@{}}Peak\\ TFLOPS\end{tabular}} & 0.13      & 0.96        & \begin{tabular}[c]{@{}c@{}}1400\\     /680\end{tabular}  & \begin{tabular}[c]{@{}c@{}}4.4(9.6)\\     /1.9(4.1)\end{tabular} \\ \hline
\textbf{\begin{tabular}[c]{@{}c@{}}Power\\ (W)\end{tabular}}    & 3.4         & 42.8        & 1,400                                                     & 4.7(7.2)                                                        \\ \hline
\textbf{Efficiency}                                             & 38.8        & 22.5        & 331.7                                                     & 406(566)                                                        \\ \hline

\end{tabular}
\begin{flushright}
\textbf{NC}: NeuroCube, \textbf{NS}: NeuroStream, \textbf{SD}: ScaleDeep, \textbf{NT}: NeuroTrainer. \\
FL: floating point, FI: fixed point, \\ FI*: fixed point with stochastic rounding\\
\textbf{Efficiency}: $GFLOPS/W$ \\
For NT, () indicates \textit{estimated} for HMC 2.0 \\
For NT, power is averaged across 8 benchmarks illustrated in Fig.~\ref{fig:summary}.

\end{flushright}
\end{table}

Table~\ref{table:prevwork} compare NeuroTrainer with previously reported DNN training accelerators. 
NeuroCube~\cite{kim2016neurocube} and NeuroStream~\cite{azarkhish2017neurostream} presents inference engines using in-memory accelerators, which can also perform training. The results demonstrate higher efficiency over a GPU-baseline showing the promise of hardware acceleration. However, performance gain is nominal as no hardware was optimized for training. 

%considerations were , demonstrates applications of 3D i shows PIM architecture for deep learning and programmability for both inference and training; however, its throughput and efficiency is too low because of on chip traffic congestion. Moreover, it assumes input duplication for fully connected layer and it is not valid in training deep network. 

%NeuroStream~\cite{azarkhish2017neurostream} uses a cluster of 32bit floating unit of PEs and with RISC core as a controller. For the computing unit, a single precision floating unit is used for both inference and training, which is not efficient in inference. With RISC controller, they provided programmability for backpropagation and weight updates but did not provide separate operation mode or data flow. Therefore, their \textit{estimated} training performance is 197 GFLOPS/s, which is only 20\% of the performance of inference.

Scaledeep~\cite{venkataramani2017scaledeep} proposes specialized hardware for different computation kernels. A multi-chip module is synthesized using five different tiles (heterogeneous architecture) and allocating layers to different tiles based on their property (such as Byte/Ops). The design demonstrates better power efficiency over GPUs. 

The main difference between NeuroTrainer and Scale- Deep is the orthogonal approaches to optimize efficiencies of different kernel. Rather than changing a data flow in the hardware for different operations as performed in~\proposedarch, ScaleDeep decides the tile distribution during design. Consequently, if the layer distribution in DNN architecture does not match the tile distributions, for example, if one kind of layer (convolution or fully connected) dominates the entire network, the tile utilization and efficiency is low. This effect is evident from~\cite{venkataramani2017scaledeep} (see Fig. 20) which shows even for various CNN benchmarks, the standard deviation of efficiency is about 28\% of average, which is expected to increase further if recurrent networks are considered. 

In contrast, the~\proposedarch~uses a homogeneous architecture and dynamically changes the data flow and data mapping to optimize the performance of individual layers. The dynamic optimization, instead of design time decisions, allow ~\proposedarch~to maintain similar throughput for much wider classes of benchmarks even including RNNs. The homogeneous architecture also makes the design easier to scale for parallel training. The secondary difference comes from the use of 3D in-memory acceleration in~\proposedarch~to reduce data movement power, and fixed point arithmetic with stochastic rounding for higher efficiency (compared to floating point in ScaleDeep).

\section{Conclusion} \label{conclusion}
We have presented~\proposedarch, an intelligent memory module with in-memory accelerators for energy-efficient training of different classes of DNNs. The ~\proposedarch~utilizes a programmable data flow based execution model to optimize memory mapping and data re-use during different phases of training operation. The simulation results demonstrate potential for appreciable performance and power-efficiency gain over baseline GPU or alternative accelerators. The ~\proposedarch~can form the building block of a scalable architecture for energy efficient training for deep neural networks. Ultimately, the performance scaling in a scalable training platform with~\proposedarch~is limited by the off-chip latency showing need for future research on better system architecture and faster off-chip network.

% A programming model and supporting architecture utilizes the flexible data flow to efficiently accelerate training of various types of DNNs. The cycle level simulation and synthesized design in 15nm FinFET shows power efficiency of ~500 GFLOPS/W, and almost similar throughput for a wide range of DNN including convolutional, recurrent, multi-layer-perceptron, and mixed (CNN+RNN) networks.

%%%%%%% -- PAPER CONTENT ENDS -- %%%%%%%%

\clearpage
%%%%%%%%% -- BIB STYLE AND FILE -- %%%%%%%%
\bibliographystyle{ieeetr}
\bibliography{ref}
%%%%%%%%%%%%%%%%%%%%%%%%%%%%%%%%%%%%

\end{document}